\documentclass[%
aip,amsmath,amssymb, reprint,
]{revtex4-1}

\usepackage{graphicx}
\usepackage{dcolumn}
\usepackage{bm}
\usepackage[ruled,vlined]{algorithm2e}
\usepackage[utf8]{inputenc}
\usepackage[T1]{fontenc}
\usepackage{mathptmx}
\usepackage{listings}
\usepackage{tikz}
\usepackage{soul}
\usetikzlibrary{arrows}
\usetikzlibrary{trees}
\usepackage{midfloat}
\usepackage{cancel}
\usetikzlibrary{decorations.pathmorphing}
\usetikzlibrary{decorations.markings}
\usetikzlibrary{automata,positioning}
\usepackage{float}
\usepackage{caption}
\usepackage{simplewick}
\usepackage{csquotes}
\usepackage{graphicx}
\usepackage{subcaption}
\usepackage{multirow}
\usepackage{amsmath}
\usepackage{cleveref}
\usepackage{tabularx}
\usepackage{natbib}
\usepackage{braket}
\usepackage{xcolor}
\definecolor{codegreen}{rgb}{0,0.6,0}
\definecolor{codegray}{rgb}{0.5,0.5,0.5}
\definecolor{codepurple}{rgb}{0.58,0,0.82}
\definecolor{backcolour}{rgb}{0.95,0.95,0.92}
\lstdefinestyle{mystyle}{
    backgroundcolor=\color{backcolour},   
    commentstyle=\color{codegreen},
    keywordstyle=\color{magenta},
    numberstyle=\tiny\color{codegray},
    stringstyle=\color{codepurple},
    basicstyle=\ttfamily\footnotesize,
    breakatwhitespace=false,         
    breaklines=true,                 
    captionpos=b,                    
    keepspaces=true,                 
    numbers=left,                    
    numbersep=5pt,                  
    showspaces=false,                
    showstringspaces=false,
    showtabs=false,                  
    tabsize=2
}

\begin{document}

\title{Operator Commutativity Screening and Progressive Operator Block Reordering toward Many-body Inspired Quantum State Preparation}
\author{Dibyendu Mondal}\affiliation{Department of Chemistry, \\ Indian Institute of Technology Bombay, \\ Powai, Mumbai 400076, India}
\author{Debaarjun Mukherjee}\affiliation{Department of Chemistry and Applied Biosciences, \\ ETH Zürich, \\ 8093 Zürich, Switzerland}
\author{Rahul Maitra}\email{rmaitra@chem.iitb.ac.in}\affiliation{Department of Chemistry, \\ Indian Institute of Technology Bombay, \\ Powai, Mumbai 400076, India} 
\affiliation{Centre for Quantum Information Computing Science \& Technology, \\ Indian Institute of Technology Bombay, \\ Powai, Mumbai 400076, India}
\begin{abstract}
In the field of quantum chemistry, the variational quantum eigensolver (VQE) has emerged as a 
highly promising approach to determine molecular energies and properties within the noisy 
intermediate-scale quantum (NISQ) era. The central challenges of this approach lie in the 
design of an expressive ansatz capable of representing the exact ground state wavefunction 
while concurrently being efficient to avoid numerical instabilities during the classical 
optimization. Owing to the constraints of current quantum hardware, the ansatz must remain 
sufficiently compact while retaining the flexibility to capture essential correlation effects. 
To address these challenges, we propose a systematic dynamic ansatz construction strategy in 
which the dominant operator blocks are initially identified
through commutativity screening, combined with an energy sorting criteria. Subsequently, the 
ansatz is progressively expanded in a stepwise manner via iterative 
operator block reordering. To minimize the overhead, the higher order 
correlation terms are incorporated via reduced lower-body tensor factorization 
in each operator block, while the adaptive construction strategy ensures that the
optimization is guided along the optimal trajectory to mitigate potential
numerical instabilities due to the presence of local traps.
Benchmark applications to various molecular systems demonstrate that this 
strategy of progressive operator-block addition achieves accurate energetics with
significantly fewer parameters while efficiently bypassing local traps. Moreover, 
in strongly correlated regions, such as bond dissociation, the method 
successfully reproduces the ground state, where other contemporary approaches 
often fail.

\end{abstract}

\maketitle
\section{Introduction}
Quantum computers, leveraging the principles of quantum entanglement and superposition, offer the 
potential to address problems that are intractable in classical computers\cite{expo,expo1}. 
However, given the constraints of Noisy Intermediate-Scale Quantum (NISQ) devices, a certain class 
of hybrid quantum-classical algorithms are preferred for molecular simulations. The Variational 
Quantum Eigensolver (VQE)\cite{Peruzzo_2014} is one of such algorithms for the 
determination of energetics of a molecular Hamiltonian. In VQE, a quantum processor is employed to construct a parameterized unitary, $U(\theta)$, to evolve a suitably chosen and problem inspired reference state ($\ket{\phi_0}$), followed by the measurement of the energy expectation value, $E(\theta)$.
\begin{equation}\label{energy}
    E(\theta)=\bra{\phi_0}U^{\dagger}(\theta)\hat{H}U(\theta)\ket{\phi_0}
\end{equation}
The energy computed by the quantum subsystem is then converted into classical data and passed to a 
classical optimization routine, which optimizes the parameters ($\theta$).
These optimized parameters are subsequently fed back to the quantum subroutine to prepare an updated 
$U(\theta)$. As a direct application of the Rayleigh-Ritz variational principle, the computed energy 
serves as an upper bound to the exact ground state (GS) energy of the molecular Hamiltonian. The accuracy 
of the results and the required gate depth in VQE are both strongly influenced by the choice of the 
parameterized quantum ansatz $U(\theta)$\cite{ucc_review,Tilly_2022}.

In the unitary coupled cluster (UCC) framework\cite{ucc_review,dUCC}, the 
GS trial wavefunction is constructed using (anti-hermitian) excitation operators.
However, while higher-rank operators are crucial for capturing strong 
correlation effects, their inclusion drastically increases circuit depth, 
rendering them impractical for current NISQ devices. The implicit inclusion of
the higher rank excitations though generalized single and double 
excitations\cite{uccgsd} also contributes to high utilization of quantum resources.
However, the construction of the exact GS wavefunction does not necessitate 
the inclusion of all possible excitation operators, since only a subset of 
determinants contributes significantly to the GS. This observation naturally 
motivates the development of a dynamic ansatz that selectively incorporates 
only the most influential operators\cite{adapt,energy_sort,DM_IJQC,sym_uccsdt}. In this context, earlier 
works by some of the present authors introduced a dynamic ansatz design 
protocol in which dominant 
operators are identified through minimal or even zero quantum measurement overhead\cite{COMPASS,compact,rbm}. 
Crucially, the higher-rank excitation effects are incorporated effectively by 
(tensor-) factoring them into a set of appropriate generalized lower-rank operators, 
often referred to as scattering operators, and its associated non-commuting 
cluster operator. This was translated to 
the formation of several operator blocks. The operator blocks were subsequently 
concatenated to develop the final ansatz-- a protocol referred to as COMmutativity Pre-screened Automated Selection of Scatterers (COMPASS)\cite{COMPASS}. Although this 
strategy yields a compact and expressive ansatz, the VQE framework still encounters 
issues with poor optimization trajectory where the landscape is swamped with 
local minima due to highly non-linear nature of the cost 
function\cite{bittel} -- a well known issue for variational optimization. This also 
translates to having the probability of observing a non-vanishing gradient along any 
given direction decreases exponentially with system size, a situation often 
described as the "narrow gorge''\cite{narrow_gorge}. This suggests that 
initializing the optimization with random starting points significantly increases 
the likelihood of the energy functional falling into these flat regions, 
known as ``barren plateaus" (BPs). Even for chemistry-inspired ansatze, it has been 
demonstrated that the energy functional 
is not immune to numerical challenges such as barren plateaus or local 
minima\cite{uccbp}. These challenges often get exacerbated with the increasing 
complexity and size of the system under consideration. Several distinct strategies 
were proposed to overcome local minima in VQE, including hybrid quantum-classical 
cost landscape modification\cite{riveradean}, collective Hamiltonian 
optimization via the snake algorithm\cite{Zhang}, unitary block optimization\cite{slattery} and judiciously constructed subspace based optimization methods\cite{chayan1,chayan2,chayan3}.

Within the UCC framework, the recently developed adaptive ansatz construction 
strategy ADAPT-VQE\cite{adapt} has emerged as a promising approach to 
overcome optimization bottlenecks such as local minima and barren plateaus 
present in the optimization landscape\cite{adapt_land}. By iteratively identifying 
the most relevant operators through a gradient-based selection procedure and 
appending them step by step, ADAPT-VQE systematically builds a compact yet 
expressive ansatz. This iterative ansatz construction has proven 
to be highly effective, leading to enhanced convergence behavior with a lesser 
probability of getting trapped in local minima. Despite its advantages, 
ADAPT-VQE faces significant challenges: its operator selection protocol relies on a gradient-based metric, where at each step the operator 
with the largest gradient is chosen for inclusion in the ansatz. While this 
procedure generally provides a very fast energy convergence route, it does 
not necessarily guarantee optimal energy reduction in every instance\cite{pruned_adapt,adapt_energy}. In 
certain cases, the selected operator may contribute only marginally to 
lowering the energy, thereby failing to stabilize the system adequately. 
More critically, in regions of the potential energy surface where the GS 
becomes nearly degenerate with low-lying excited states, the pool of 
operator gradients may decay exponentially, a phenomenon referred to 
as a 'gradient trough'. Under such conditions, ADAPT-VQE can get stuck 
during the operator selection process, severely 
hindering its ability to converge to the correct ground state.

To address these challenges, in this work, we introduce a heuristic strategy in which
the dominant operator blocks are first constructed through a commutativity-based 
screening, followed by the block selection based on an energy-sorting procedure. 
Importantly, this requires only a minimal number of quantum measurements. 
Each operator block 
consists either of an important two-body excitation operator along 
with a set of scatterers, which are capable of generating higher-rank 
excitation effects, or alternatively, one single-excitation operator. 
These operator blocks collectively constitute the operator block pool to 
choose from while the ansatz is constructed dynamically in a progressive 
manner. Instead of the gradient based selection of individual operators, 
the operator blocks are dynamically chosen at each optimization step through a 
local VQE micro-cycle that guides the optimization along the steepest pathway 
having the maximum energy stabilization. This block wise and stability-driven 
construction protocol is referred to as COMPASS with Progressive block 
ReOrdering (COMPASS-PRO)-- a more accurate and highly robust
variant of COMPASS\cite{COMPASS}.

The essential advantages of COMPASS-PRO lies in two key features. First, the 
incorporation of scatterers allows the ansatz to effectively capture higher-order
excitation effects with only lower-rank operators, which significantly reduces 
the quantum gate count. Second, the progressive stabilization-driven selection 
of operator blocks provides flexibility in the ansatz growth, thereby 
avoiding local numerical traps and ensuring a smooth optimization trajectory 
toward the exact GS. Through simulations of systems with moderate to strong 
correlation, we demonstrate that COMPASS-PRO generates compact yet very expressive 
ansatz than the existing ansatze, without compromising energy accuracy, 
and successfully avoids numerical issues during classical optimization. Most 
notably, in regions of the potential energy profile where the GS becomes nearly 
degenerate with low-lying excited states, COMPASS-PRO demonstrates a 
distinct advantage: it successfully converges to the true ground state by 
following a more favorable optimization path, whereas state-of-the-art methods 
such as ADAPT-VQE often fail due to gradient trough.


\section{Theory}
\subsection{Choice of Operators in UCC Ansatz}
As highlighted in the introduction, the unitary coupled-cluster (UCC) ansatz restricted to 
single and double excitation operators is insufficient to fully capture correlation energy in 
strongly correlated molecular regimes. However, directly incorporating higher-order excitation 
operators leads to a substantial increase in circuit depth, which exceeds the practical limits 
of near-term quantum devices.

An efficient alternative is to generate higher-order correlation effects indirectly through 
the inclusion of generalized two-body operators within the ansatz. Among the different classes 
of two-body generalized operators, those having only one quasi-orbital destruction operator are 
particularly effective. This specific class of generalized operators is 
referred to as scattering operators or scatterers ($S$). Depending on whether the orbital destruction operators in scatterers correspond to hole or particle types, 
they are denoted as, $S_h$ and
$S_p$, respectively, $S=S_h + S_p$\cite{iccsdn,uiccsdn,intro_s}.
\begin{eqnarray}
    S_h = \frac{1}{2} \theta_{ij}^{am}a_{a}^{\dagger}a_m^{\dagger}a_j a_i \nonumber \\
    S_p = \frac{1}{2} \theta_{ie}^{ab}a_{a}^{\dagger}a_b^{\dagger}a_e a_i
\end{eqnarray}
Here $i,j,...$ and $a,b,...$ denote the occupied and virtual spinorbital indices, respectively with respect to the Hartree-Fock (HF) vacuum. The quasi-hole/particle  destruction operators (denoted as $m$ and $e$ respectively) in one of the scatterer vertices are restricted to a set of \textit{active} occupied (hole) and unoccupied (particle) spinorbitals and they together form a contractible set of orbitals (CSOs).
Depending on the commonality of CSOs, when such scatterers act 
following a two-body excitation operator($T_{ij}^{ab}$), they can induce triple excitation effects (via an effective $T_{ijk}^{abc}$).  
\begin{equation}\label{triples}
   \sum_{m} S_{ij}^{am}T_{mk}^{bc}\xrightarrow{} T_{ijk}^{abc} \hspace{0.2cm};  
   \sum_{e} S_{ie}^{ab}T_{jk}^{ec}\xrightarrow{} T_{ijk}^{abc}
\end{equation}
As the scatterers have one effective hole-particle excitation structure, 
each such contraction with $N$-body excitation operator generates one rank 
higher ($(N+1)$-body) excitation effects: 
${\contraction{}{{S}}{}{{T_2}}
{S} {T_2}}\rightarrow {T_3}$,
$\contraction[2ex]{}{{S}}{}{\contraction{}{{S}}{}{{T}_{2}} {S} {T}_2} {S} \contraction{}{{S}}{}{{T}_{2}} {S} {T}_{2}
\rightarrow {T}_4...$.
On quantum computers, the higher order effects are generated through the nested commutators of the corresponding anti-hermitian operators ($[\sigma,\tau_2] \longrightarrow 
\tau_3$ or $[\sigma,[\sigma,\tau_2]] \longrightarrow \tau_4$) which appear in
a factorized (disentangled) UCC ansatz\cite{dUCC,karol1}. 
\begin{equation}
    e^{\sigma}e^{\tau} = e^{\sigma +\tau +[\sigma,\tau]+[\sigma,[\sigma,\tau]]+...}
\end{equation}
Here $\tau$ and $\sigma$ represent
the anti-hermitian form of the cluster and scattering operators, respectively. 
Owing to the presence of a quasi-orbital destruction operator, the action of a scatterer ($S$) results 
in annihilation of the reference HF state leading to the vacuum annihilating condition (VAC), 
$S\ket{\phi_0}=0$ (also consequently, $\sigma\ket{\phi_0}=0$).
It is important to note that due to the VAC, not all excitation and 
scattering operators contribute significantly to the construction of the ground-state 
wavefunction, and the overall accuracy is strongly influenced by the ordering of the dominant excitation and scattering operators. Therefore, a well-defined strategy is essential for constructing an effective ansatz that systematically pairs up appropriate excitation and scattering operators, leveraging their non-commutativity 
to induce higher rank excitations while retaining its trainability in NISQ architecture.

\subsection{Construction of Operator Block Snippets and a Road-map toward a Dynamic Ansatz Design} \label{compass}

The careful selection of dominant excitation and scattering operators plays a crucial role in 
designing a compact yet expressive ansatz. 
We first select the most dominant double excitation 
operators by optimizing single
parameter circuits $E_I = \bra{\phi_{HF}}e^{-\tau_I(\theta_I)}He^{\tau_I(\theta_I)}\ket{\phi_{HF}}$.
Here $\tau_I=T_I-T_I^{\dagger}$, and $I$ represents the combined hole-particle spinorbital indices of
two-body excitation operator.
Only those $\tau_I$'s are selected for which $\Delta E_I=|E_I-E_{HF}|$ is greater than
a predefined threshold and is placed in what will be termed an operator block. The operator blocks will be denoted by $(\alpha, \beta, \gamma, \delta...)$. 

\begin{equation}
    U_{\alpha}=\Big[e^{\tau_I(\theta_I)}\Big]_{\alpha}
\end{equation}
Note that till now each operator block is restricted to the leading two-body 
contributions, however, their relative ordering remains undefined. In the regime of strong correlation, it becomes essential to incorporate 
higher-body excitation effects as well. These higher-body effects are systematically included by 
extending the constructed operator blocks through the incorporation of 
suitable two-body scattering operators. As mentioned before, the generation of 
higher-order excitations is governed by the commonality of CSOs, or more precisely,
by the non-commutativity between excitation and scattering operators in each block.
Consequently, not every scatterer can generate higher-order excitation 
effects when combined with arbitrary excitation operators. Another equally important point is 
that not all of the generated higher-order excitations contribute significantly to the ground-
state energy. In order to identify the suitable scatterers within an operator block that can 
induce dominant higher-order effects, we have applied a two-step selection procedure. 
First, appropriate scattering operators are chosen 
based on their non-commutativity with the excitation operator present in the 
operator block, ensuring the generation of higher-order excitation manifolds. 
Subsequently, only those pair of operators (cluster operator and its non-commutative scatterer pair) are kept in a block for which the resultant higher body excitation is
dominant. This is ascertained by the relative stabilization of the two-parameter energy functional:
$E_{I\mu} = \bra{\phi_{HF}}\overline{e^{-\tau_I(\theta_I)}e^{-\sigma_{\mu}(\theta_{\mu})}}H\overline{e^{\sigma_{\mu}
(\theta_{\mu})}e^{\tau_I(\theta_I)}}\ket{\phi_{HF}}$.
Here $\sigma_{\mu}=S_{\mu}-S_{\mu}^{\dagger}$ and $\mu$ represent the combined spinorbital indices of a scatterer and the overline indicates there is a commonality of CSO in $\tau_I$ and $\sigma_{\mu}$ which 
results a higher order excitation as in Eqn. \ref{triples} . 
Finally, we add those scattering
operators in that particular block for which $|E_{I\mu}-E_I|$ is greater than a predefined 
threshold. In this manner each operator block is expanded. 
 
\begin{equation}
    U_{\alpha}=\Big[\prod_{\mu}\overline{e^{\sigma_{\mu}(\theta_{\mu})}e^{\tau_I(\theta_I)}}\Big]_{\alpha}
\end{equation}
With the construction of various operator blocks containing one two-body excitation operators
and a series of non-commuting scatterers, along with all the single excitation operators, 
one may simply concatenate them in terms of their relative stabilization. This leads to a disentangled ansatz of the form:
\begin{equation}\label{final_compass}
    \ket{\psi(\theta)}=\prod_s e^{\tau_s(\theta_s)} \prod_\alpha U_{\alpha}\ket{\phi_{HF}}
\end{equation}
The ordering among various operator blocks, $\alpha$, plays a crucial role in
accuracy and the convergence landscape of the ansatz.
Although the present authors had previously proposed the idea\cite{COMPASS} in which the operator blocks were formed based on the commutativity between
the cluster operators and the scatterers, and were further energetically ordered, the operator blocks in the resulting ansatz were
far from optimally ordered. This often results in getting trapped in one of the local minima during their variational optimization, and without an effective "ansatz growth and landscape burrowing" mechanism, the result may
sometimes be inaccurate. There remains ample opportunity to adopt a different ansatz engineering strategy towards significantly shallower structure that fits NISQ hardware.
In the current work, we heuristically demonstrate that it is possible to prepare an
extremely shallow anastz structure by adaptive addition of the operator blocks that 
generates highly accurate energy across the molecular potential energy profile with minimal
quantum resources. Furthermore, such a strategy is shown to bypass the local traps and
BPs more efficiently than the existing dynamic ansatze, albeit with somewhat higher 
measurement overhead.

\subsection{COMPASS with Progressive block Re-Ordering (COMPASS-PRO)}
\subsubsection{Compression of Two-body Operator Block through Removal of Redundant Pathways:}\label{redundancy step}

The operator blocks containing one and two-body operators that are generated via
commutativity screening are tailored to capture the correlation energy effectively. However, there may exist multiple non-commutative pathways (with several pairs of 
double excitation operators and scatterers) that yield the same dominant 
three-body operator ($\tau_X$). Here, $X$ represents the combined hole–particle 
indices of the corresponding three-body operator. For example, a dominant 
three-body operator $\tau_X$ is generated through two or more possible combinations 
of two-body excitation and scattering operators, e.g., $[\sigma_{\mu},\tau_I] \longrightarrow 
\tau_X$ or $[\sigma_{\nu},\tau_J] \longrightarrow \tau_X$, and so on. Note that the
various scatterer-excitation operator pairs $(\sigma_{\mu},\tau_I)$, 
$(\sigma_{\nu},\tau_J)$, ... necessarily belong to two different operator blocks 
(as each operator block accommodates only one cluster operator), 
both of which got selected via the thresholding criteria over one and two 
parameter operator blocks. This leads to redundant 
occurrences of identical triple-excitation contributions. Since such redundant 
triple-excitations do not contribute towards additional correlation, it is imperative that 
such a redundancy must be systematically eliminated by symbolical identification 
to reduce the resource count.

Given that we have previously computed $E_I$ for all two-body excitations
and $E_{I\mu}$ for all allowed $(\sigma_{\mu},\tau_I)$ non-commuting pairs
(some of which are effectively redundant), one may break the $(\sigma_{\mu},\tau_I)$
pairing in the corresponding operator blocks and retain only that particular 
pair which has the highest stabilization contribution. This implies that only 
that particular $(\sigma_{\mu},\tau_I)$ pair is retained for which the energy 
difference $\Delta E_{I\mu} = |E_{I\mu} - E_I|$ is maximum. This is to retain 
the most dominant pathway toward the generation of higher excitation terms.
It is very important to note here that the breaking of the $(\sigma_{\mu},\tau_I)$ pair is 
achieved by discarding only the $\sigma_{\mu}$’s from the corresponding 
operator block, while the $\tau_I$ is retained in the block.

The removal of redundant operators results in a refined collection of
two-body operator blocks, denoted $U_{\alpha}$, where each block contains the same 
number or fewer scatterers than those originally obtained by commutativity screening 
(see Sec. \ref{compass}). Alongside these refined two-body blocks, we 
also retain all one-body operators, each containing a single-excitation 
operator.

\subsubsection{Progressive Ansatz Construction via Operator Block Reordering and Step-wise Optimization}

The removal of redundant pathways discussed in the previous section results in 
a collection of operator blocks, $U_\alpha$, each of which contains one cluster 
operator and zero, one, two,...scatterers. It is now an open question how these
operator blocks would be ordered which critically decides the accuracy and 
the optimization landscape. At this point, one could in principle optimize 
over all the selected operator blocks (ordered in a particular way) to obtain 
the ground-state wavefunction. However, in often cases for global ansatz 
optimization, not all operator blocks which are chosen through individual
parametrized block optimization, are essential or contributing toward 
effective energy minimization. Furthermore, the accuracy may be dictated 
by the block ordering. As the system size increases, the number of operator 
blocks also grows (as the number of cluster operator), which often results 
in rugged optimization landscape swamped with local traps. 
Here we choose to build up a complete ansatz adaptively by adding 
one operator block at each step such that one may achieve tunable 
accuracy while concurrently ensuring a guaranteed systematic 
burrowing of the optimization landscape toward exact GS energy.

With the chosen set of blocks in the operator block pool, we build up the
ansatz, guided by a local energy minimization and sorting.
Starting with the HF reference state, one may variationally 
optimize with each of the chosen operator blocks and select the particular 
operator block that stabilizes the reference most. The parameters in 
the chosen operator block is fixed at their optimized values and the HF 
reference is rotated to a new reference. This process continues, at each step
with a new set of reference, until the convergence is achieved. This implies 
that at this stage, no additional operator block can significantly improve 
the reference state and the wavefunction at this stage is taken to be the 
GS wavefunction.

The overall structure of this algorithm is as follows:
\begin{itemize}
    \item \textbf{Initialization:} Initialize the qubits to an appropriate reference state. Here we choose HF state as the reference wavefunction in the first step.
    \item \textbf{Operator Block Formation:} Keeping HF state as reference, dominant two-body excitation operators are identified through a one-parameter energy optimization scheme, and the selected operators are placed in distinct operator blocks. Each such block is subsequently expanded by incorporating relevant scatterers, determined via non-commutativity with the previously selected cluster operators and refined through a two-parameter optimization strategy, as described in Sec. \ref{compass}. 
    
    \item \textbf{Block Refinement through Removal of Redundant Pathways:} Every operator block undergoes a refinement step where various different pathways leading to redundant higher-order determinants are systematically eliminated, details of which are provided in Sec. \ref{redundancy step}. In addition to these refined two-body operator blocks, all the one-body operators are also assembled into separate operator blocks.
    
    \item \textbf{Local VQE Micro-Cycles:} On top the reference, we prepare an ensemble of trial states by acting 
    all the operator blocks and perform local VQE optimization (with already added parameters frozen to their optimized values at the previous step).
    If multiple quantum devices are available, the optimization with various operator blocks can be performed in parallel. 
    \begin{equation}
        E_{\alpha}^{(k)}=\bra{\phi_0^{(k-1)}}U^{\dagger}_{\alpha}HU_{\alpha}\ket{\phi_0^{(k-1)}}
    \end{equation}
    Here, $\ket{\phi_0^{(k-1)}}$ is the optimized reference state obtained in the previous $(k-1)$-th step. (In case of 1st step, reference state is the HF state $\ket{\phi_0}$.)
   \item \textbf{Operator Block Selection and Ansatz Growth:} Choose the operator block which has the highest energy stabilization ($E_{\alpha}^{(k)}$). This implies that the optimization is channelized to
   the steepest burrowing direction at this stage. The operator block ($U_{\alpha}$) is added to the existing ansatz. Thus, in the $k$-th step, the final ansatz becomes:
    \begin{equation}
        \ket{\psi^{(k)}}=U_{\alpha}\ket{\phi_0^{(k-1)}}
    \end{equation}
    \item \textbf{Global VQE Macro-Cycle:} Perform a VQE optimization, refining all parameters. Use the previously optimized parameters as the starting point. The final optimized energy in the $k$-th step is given by:
    \begin{equation}
        E^{(k)} = \bra{\phi_{HF}}\underbrace{U^{\dagger}_{\gamma}...U^{\dagger}_{\beta}...U^{\dagger}_{\alpha}}_{\substack{\text{k-number of}\\ {\text{operator blocks}}}}H
        \underbrace{U_{\alpha}...U_{\beta}...U_{\gamma}}_{\substack{\text{k-number of}\\ {\text{operator blocks}}}}\ket{\phi_{HF}}
    \end{equation}
    \item \textbf{Progressive Reference Update:} If the energy difference between two successive cycles ($\Delta E=|E^{(k)}-E^{(k-1)}|$) is greater than a pre-defined threshold, return to Local VQE Micro-cycles step. The optimized wavefunction from the $k$-th step serves as the reference for the $(k+1)$-th step, and $E^{(k)}$ is the new optimized reference energy.
    \begin{equation}
        \ket{\phi_0^{(k)}}=\underbrace{U_{\alpha}...U_{\beta}...U_{\gamma}}_{\substack{\text{k-number of}\\ {\text{operator blocks}}}}\ket{\phi_{HF}}
    \end{equation}
\end{itemize}
The benefit of this algorithm lies in its stepwise optimization of energy where
each operator block is chosen through local optimization and is updated through global optimization at each macro-cycle, ensuring its guaranteed robustness to navigate past any local trap at any given optimization cycle.

\begin{figure*}[t]
    \centering
    \includegraphics[width= 18cm, height=12cm]{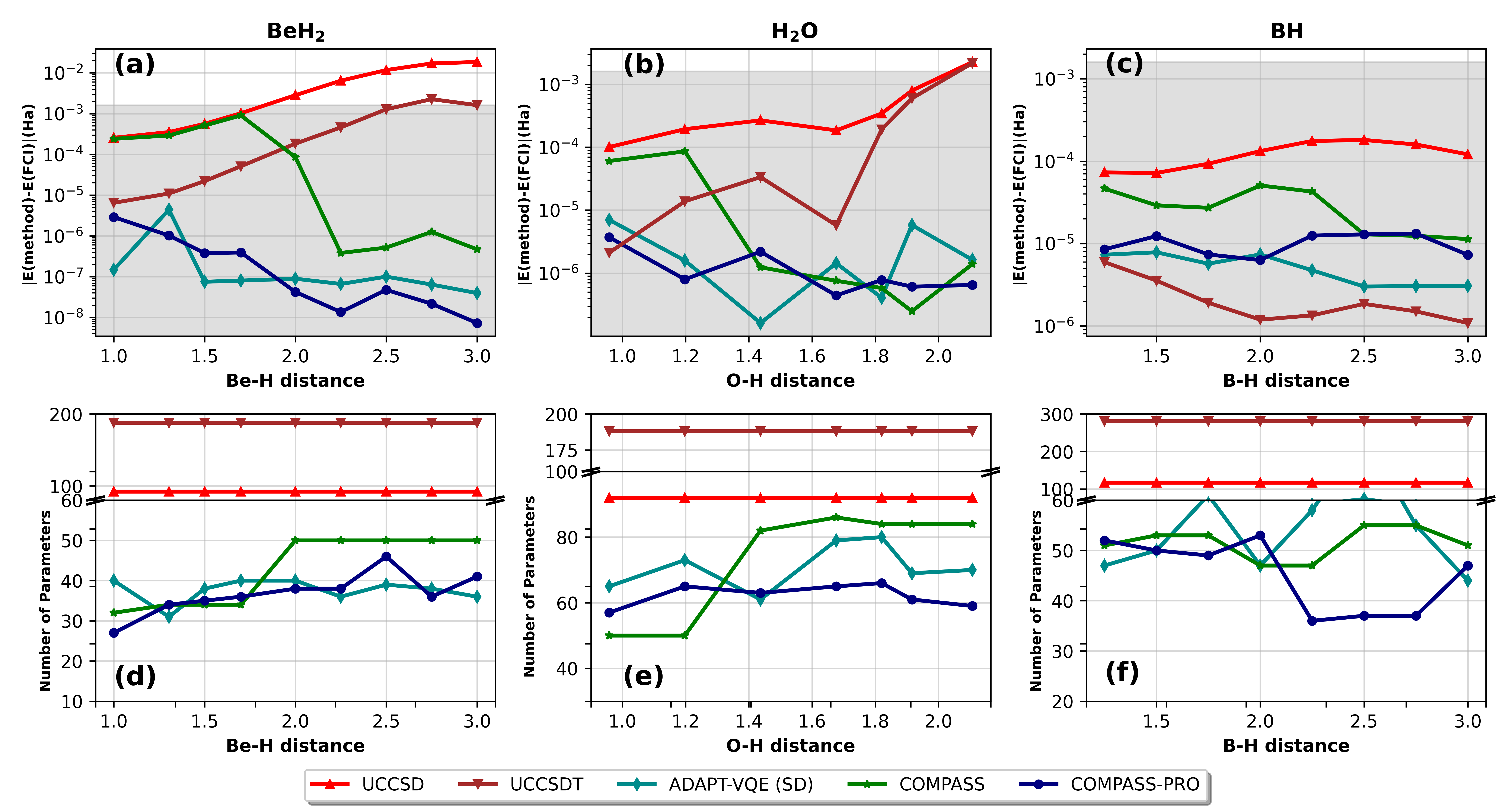}
    \caption{\textbf{Accuracy as a function of the bond length parameter for UCCSD, UCCSDT, COMPASS, ADAPT-VQE (SD) and COMPASS-PRO is plotted with respect to FCI: (a) linear $BeH_2$, (b) $H_2O$, and (c) $BH$. The shaded region indicates chemical accuracy. Sub-figures (d), (e) and (f) estimate parameter counts for $BeH_2$, $H_2O$ and $BH$, respectively.}}
    \label{fig:acc vs params}
\end{figure*}

\section{Results}
\subsection{General Numerical Considerations}
All the calculations were carried out using Qiskit-Nature\cite{qiskit_nature} which 
imports the one- and two-body integrals from PySCF\cite{pyscf}. The STO-3G basis
set\cite{sto} is employed for all molecular simulations, along with a direct 
spin-orbital to qubit mapping. To map the second quantized fermionic operators to 
qubit operators, we applied the Jordan-Wigner encoding\cite{JW,JWT}. In the 
classical optimization phase, we employed the L-BFGS-B optimizer\cite{B,F,G,S} to 
minimize the energy functional. To reduce the quantum resource requirements, 
we have taken only those
scattering operators for which the excitation vertex and scattering vertex 
belong to two different spin sectors, along with the restrictions on CSOs 
that are allowed to span only the HOMO and LUMO for all the test cases. In all test case 
simulations, during the operator block construction step, we include two-body excitation 
operators with $\Delta E_I > 10^{-5}$ and scatterers satisfying $\Delta E_{I\mu} > 10^{-6}$. 
Operator blocks are successively added to the COMPASS-PRO ansatz until the energy difference 
($\Delta E$) between two consecutive Global VQE macro-cycles falls below $10^{-7}$.

\subsection{Molecular Potential Energy Profile and Parameter Count:}
In this section, we present a comparative analysis of the energy accuracy 
with respect to Full Configuration Interaction (FCI) and the associated
parameters count for several ansatze: 
UCCSD, UCCSDT, COMPASS, ADAPT-VQE with the SD operator pool, and the ansatz 
generated by COMPASS-PRO. We focus on three challenging test cases: 
symmetric $Be-H$ bond stretching in linear $BeH_2$, symmetric $O-H$ bond 
stretching in $H_2O$ and single bond stretching of $BH$. 
The core $1s$ orbital of $Be$ and $O$ for $BeH_2$ and $H_2O$, respectively, were
kept frozen throughout. As shown in Fig.\ref{fig:acc vs params},
Expectedly, COMPASS-PRO consistently achieves higher accuracy across all the molecular
the potential energy surfaces, compared to both UCCSD and UCCSDT ansatze. 
Moreover, in nearly all cases, COMPASS-PRO surpasses the performance of COMPASS while 
utilizing substantially fewer variational parameters. This directly implies a reduced 
quantum gate count in COMPASS-PRO, as the number of variational parameters corresponds 
to the total number of operators included in the ansatz.
Although the ADAPT-VQE method
sometimes achieves a level of accuracy comparable to that of COMPASS-PRO, 
it generally requires a higher number of parameters to reach similar 
accuracy. These results highlight the superior efficiency and 
expressive power of COMPASS-PRO, particularly under resource-constrained 
quantum simulations.

\begin{figure*}[t]
    \centering
    \includegraphics[width= 14cm, height=14cm]{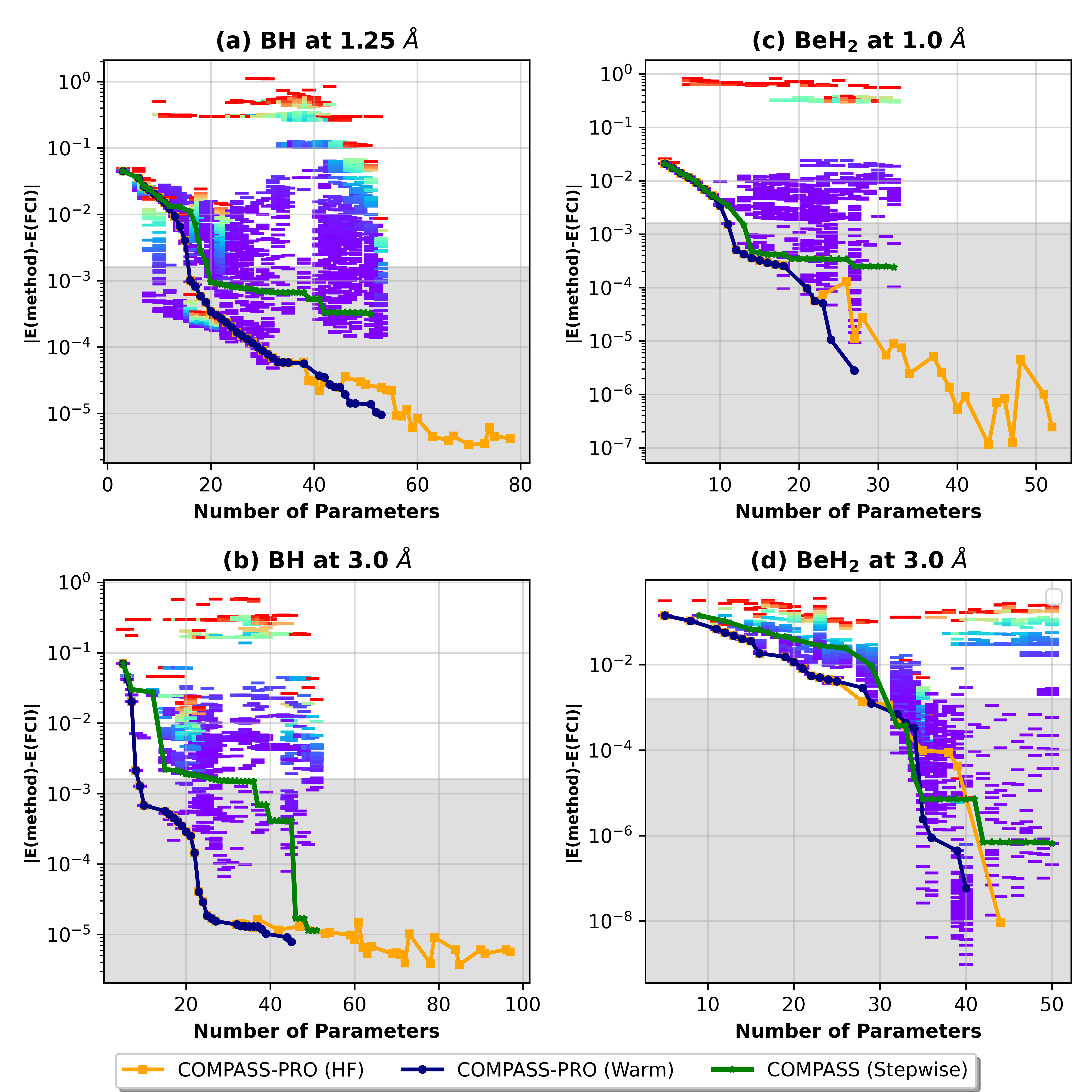}
    \caption{\textbf{Energy error across successive macro-cycles is shown for COMPASS-PRO (warm), COMPASS-PRO (HF), and COMPASS (stepwise) ansatze. The dashed lines represent the optimized energy error achieved at each intermediate ansatz length for COMPASS-PRO (warm) and COMPASS (stepwise), where the parameters are initialized from random starting values at the global VQE macro-cycle.}}
    \label{fig:loc_minima}
\end{figure*}

\begin{figure*}[t]
    \centering
    \includegraphics[width= 16cm, height=9cm]{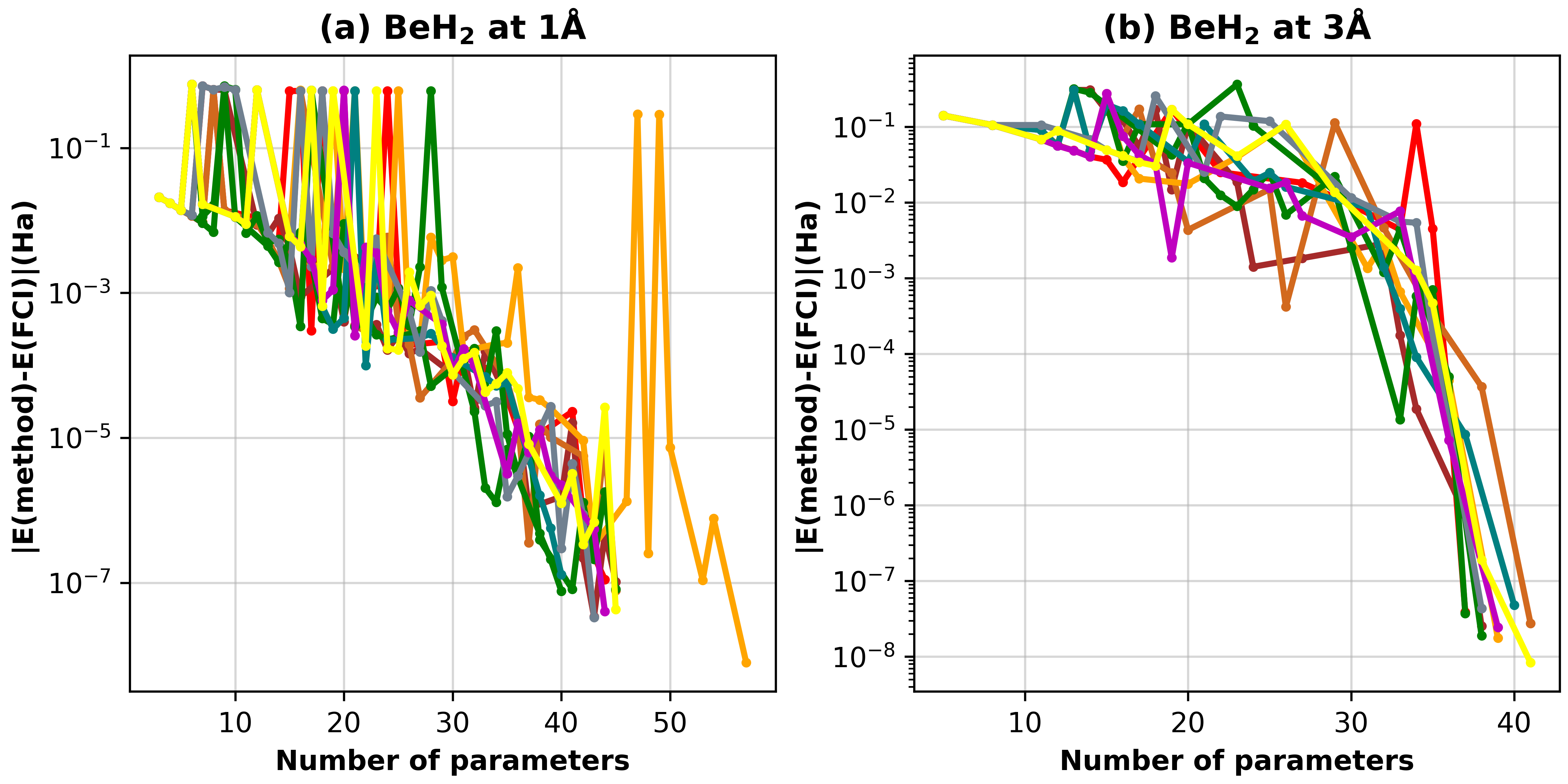}
    \caption{\textbf{Energy accuracy with respect to FCI for 10 independent runs is plotted at each macro-cycle of COMPASS-PRO for two distinct geometries of linear $BeH_2$.
    In each macro-cycle, the global VQE optimization is performed, initialized from a randomly chosen set of parameter values. While the random initialization may occasionally trap the intermediate ansatz in local minima, the incorporation of the most stabilizing operator block in the subsequent cycle deepens these minima and guides the optimization along a quasi-optimal pathway toward the global minimum.}}
    \label{fig:traps}
\end{figure*}
\subsection{Optimization Landscape and Distribution of Local Minima:}
In this section, we present a systematic numerical analysis of the 
occurrence and distribution of local minima within the energy landscape as 
the COMPASS-PRO ansatz is adaptively expanded. 

\textit{(i)} We begin with the standard execution of the COMPASS-PRO protocol, where
each step of global VQE optimization is initialized from 
the previously converged parameter set. This procedure is referred to as COMPASS-PRO 
(Warm) in Fig.\ref{fig:loc_minima}. 

\textit{(ii)} At each ansatz length of COMPASS-PRO (warm) as described above, 
we employ the (intermediate) ansatz to explore local minima by global macro-cycle 
optimization of the energy functional starting from a set of randomly selected 
initial parameters. That implies at each $k^{th}$ step, all the parameters included
till that point is randomly initialized within the interval of $2\pi$ 
(from $-\pi$ to $\pi$). A set of 50 such experiments with random initializations 
during each global optimization phase of the COMPASS-PRO algorithm was carried out, 
and the corresponding optimized energies are recorded which represent the corresponding 
local minima within the parameter landscape. These are demonstrated as discrete color
codes (red to violet, arranged in the order of descending energy error).

\textit{(iii)} In addition to this, we also investigate the performance of the COMPASS-PRO 
ansatz under a distinct initialization scheme, where all variational parameters in the 
global VQE macro-cycle are initialized to zero. This effectively corresponds to 
starting from the Hartree–Fock reference state at each stage, and the results 
obtained from this protocol are denoted as COMPASS-PRO (HF) in Fig.\ref{fig:loc_minima}.

\textit{(iv)} Alongside COMPASS-PRO, we also explore the energy landscape associated with 
the COMPASS ansatz, where now the circuit is incrementally constructed 
by appending one operator block at a time. The ordering of the blocks added one after 
the other is dictated by the amount of stabilization induced by the constituent 
two-body cluster operator present in the corresponding block. Here, 
global VQE optimization begins from the optimal set of parameters of the 
preceding stage (warm start), and to study the existence of local minima, 
we have additionally performed multiple optimizations with 50 random initial 
parameters sampled within $[-\pi, \pi]$ at every macro-cycle as above.

The above numerical experiments are carried out for two geometries of the linear 
$BH$ molecule and two geometries of the linear $BeH_2$ molecule and the observations 
are explained below. \\

\textbf{$BH$ molecule:} 
In Fig.\ref{fig:loc_minima} (a) and (b), we have shown the variation of the energy error with respect to the FCI 
as a function of parameter count across successive macro-cycles for $B-
H$ at bond lengths 1.25\AA and 3\AA respectively. From the figure it is evident that COMPASS-PRO (HF) requires a substantially larger number 
of parameters to achieve the same level of accuracy as COMPASS-PRO (Warm).
In almost all the instances, when global VQE optimization is carried out from a randomly 
initialized parameter set at each macro-cycle, the resulting optimized energy remains 
much higher than that obtained from COMPASS-PRO (Warm). These energy values correspond 
to the trapping in local minima at that ansatz length, whereas COMPASS-PRO (Warm) 
consistently bypasses such unfavorable regions of the optimization landscape.
Another key observation is that for small ansatz sizes, the spread of energies 
obtained from random initializations remains relatively narrow. However, as the ansatz 
size grows, the range of energies expands significantly, reflecting the increasingly 
rugged optimization landscape. However, at nearly all ansatz depths, the accuracy of 
COMPASS-PRO (Warm) consistently exceeds those obtained from random 
initializations. These findings strongly support the idea that a well-informed 
initialization strategy outperforms both random initialization and HF initialization.
Finally, as expected, the convergence of COMPASS-PRO (Warm) with respect to parameter 
count is markedly faster than that of the COMPASS (Stepwise) ansatz. To achieve a 
given accuracy threshold, COMPASS-PRO (Warm) requires far fewer parameters than 
COMPASS (Stepwise), highlighting the efficiency of the operator block reordering strategy.

\textbf{$BeH_2$ molecule:} 
A similar behavior is observed for the $BeH_2$ system at $Be-H$ bond lengths of 1\AA and 
3\AA. As illustrated in Fig.\ref{fig:loc_minima} (c) and (d), COMPASS-PRO (Warm) consistently demonstrates superior 
performance over COMPASS-PRO (HF). 
In particular, at the $Be-H$ bond length of 1\AA, COMPASS-
PRO (HF) demands a significantly larger number of parameters compared to COMPASS-PRO (Warm) 
to achieve a comparable level of accuracy, and moreover, its optimization trajectory does not 
follow a smooth monotonic convergence pattern. Analogous to the $BH$ case, the energies obtained 
through global VQE optimization at each macro-cycle, when initialized with a random parameter 
set, remain significantly higher than those from COMPASS-PRO (Warm). This again indicates 
that COMPASS-PRO (Warm) is capable of systematically avoiding local minima present in the 
optimization landscape at every stage. Furthermore, similar to $BH$, the number of local 
minima proliferates as the ansatz size increases. As expected, the convergence behavior of 
the COMPASS (Stepwise) ansatz is also much slower compared to COMPASS-PRO (Warm), which 
achieves the target accuracy with far fewer parameters.

\begin{figure*}[t]
    \centering
    \includegraphics[width= 18cm, height=13cm]{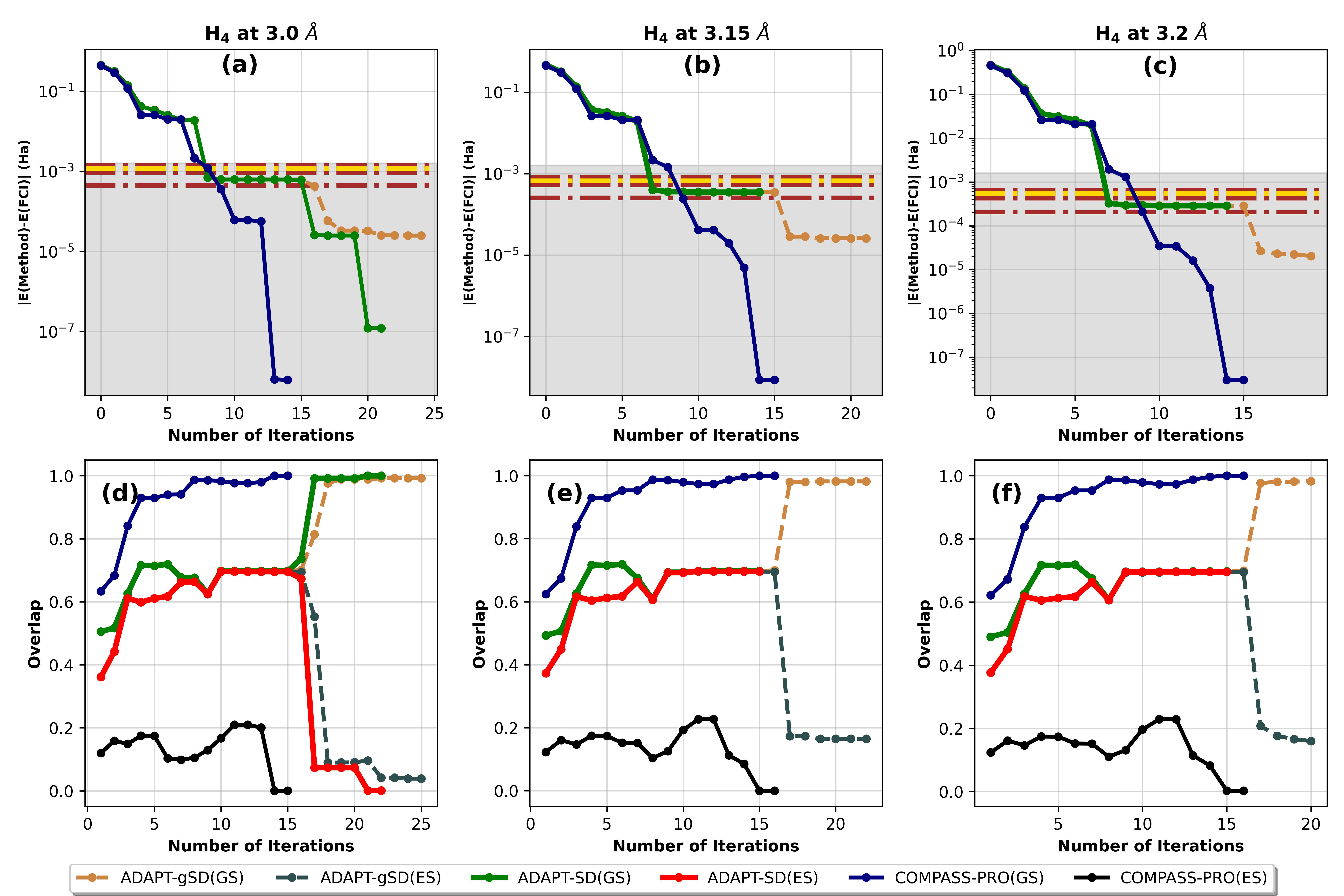}
    \caption{\textbf{In (a), (b) and (c) accuracy with respect to FCI is depicted for three geometries of linear $H_4$ at each macro-cycle of COMPASS-PRO, ADAPT-VQE (SD) and ADAPT-VQE (gSD).  In (a), (b) and (c) the horizontal brown and yellow dotted-dashed lines indicate the energy difference with FCI GS for several low-lying FCI ES. The yellow color is for first $^1A_{1g}$ ES. Plot (d), (e) and (f) show the corresponding overlap convergence of COMPASS-PRO, ADAPT-VQE (SD) and ADAPT-VQE (gSD) with $^1A_{1g}$ ground state (GS) and first $^1A_{1g}$ excited state (ES).}.}
    \label{fig:overlap}
\end{figure*}

\subsection{Progressive Ansatz Construction and Optimization with Random Initialization: The 
Burrowing Mechanism}
In the previous section, we investigated the positions of local minima where the 
parameters were randomly initialized at every intermediate step (random 
initialization for global VQE macro-cycle), but the ansatz was
constructed through COMPASS-PRO (Warm). It was observed that COMPASS-PRO, when 
warm-initialized (COMPASS-PRO (warm)) throughout (for all global VQE macro-cycles) 
was able to bypass all of the local traps at every ansatz depth. We now turn our 
attention to the scenario where the COMPASS-PRO ansatz gets trapped in a 
local minimum during a particular macro-cycle (with random initialization) and 
investigate whether it can subsequently escape from the local trap and progress 
toward the global minimum. To address this, we performed a study in which, 
at each global VQE macro-cycle, the optimization was initiated from a randomly 
chosen parameter set to let the optimization get stuck in a local minimum. 
The corresponding "optimized" parameter values (of the previous step) were then 
taken forward into the subsequent local VQE micro-cycle to select the optimal 
operator block. In this setup, our block selection criteria via energy 
stabilization ensures that the chosen operator block is the most dominant one 
for that step with the 
greatest potential to pull the subsequent optimization out of the local trap.
For this experiment, we considered the $BeH_2$ molecule at two bond lengths, 
1\AA and 3\AA, and independently repeated the procedure 10 times for each geometry. 
As shown in Fig.\ref{fig:traps} , the convergence of the COMPASS-PRO ansatz is no longer strictly 
monotonic but instead becomes frequently trapped in several local minima scattered 
throughout the optimization landscape. However, quite remarkably, each time the 
optimization is stuck in a local minimum at a given macro-cycle, the inclusion 
of the next operator block enables the ansatz to escape from that trap in the 
subsequent cycle, eventually leading to highly accurate energies with respect 
to FCI for all the cases for both the geometries.

Quite convincingly, the COMPASS-PRO algorithm possesses a in-built mechanism to 
overcome local traps during the optimization process through alternate local 
micro-cycle-based growth and global macro-cycle-based optimization. This 
robustness further arises from the adaptive construction of the ansatz, where 
the addition of the most optimal operator block at each stage 
deepens the local minima and provides a quasi-optimal path towards 
reaching the global minimum.

\subsection{Adaptive Operator Block Augmentation vs Gradient Based Ansatz Construction: a Comparative Study of Energy Accuracy and Wavefunction Overlap}

In this section, we present a comparison of the energy and overlap convergence 
of the COMPASS-PRO 
(Warm) ansatz against ADAPT-VQE (SD) and ADAPT-VQE (gSD), evaluated step by step during the 
ansatz construction process. Such a comparison is crucial for assessing whether such heuristic ansatz design 
strategies are capable of consistently converging to the exact ground 
state across the full potential energy surface of a molecule. This becomes particularly important 
in scenarios where certain excited states lie below the HF reference state and 
close to the true ground state energy. In such cases, starting from the HF reference, these ansatz 
design protocols may fail to recover the correct ground state and often gets trapped in a 
local minima around a low lying excited eigenroot.

We have demonstrated this in the bond dissociation regime of the 
linear $H_4$ system. To analyze the behavior in detail, we computed the exact 
FCI ground state and 
several low-lying excited states (shown as dashed lines in Fig.\ref{fig:overlap} ) at three distinct $H-H$ bond 
lengths. Across all geometries, the ground state belongs to the $^1A_{1g}$ 
irreducible representation. Using the HF reference as a starting point, we 
constructed the ground-state wavefunction using both the ADAPT-VQE and COMPASS-PRO-VQE 
methods. At each macro-cycle during ansatz construction, we monitored the energy error 
with respect to the FCI ground state, along with the overlap of the intermediate ansatze 
(after every macro-cycle) with both the ground and first excited $^1A_{1g}$ (yellow line 
in Fig.\ref{fig:overlap}) FCI states.

From the Fig.\ref{fig:overlap}, it is evident that at $H-H = 3$\AA, ADAPT-VQE with the SD 
operator pool initially reaches a region where the overlaps with both the ground and first 
excited $^1A_{1g}$ states are nearly identical. Interestingly, after continued operator 
additions, it eventually converges to the correct ground state, characterized by nearly 
unit overlap with the ground state and nearly vanishing overlap with the excited state. 
However, at slightly larger bond distances
(3.15\AA and 3.2\AA), ADAPT-VQE (SD) becomes trapped in a region where the overlaps 
with the ground and first excited $^1A_{1g}$ states remain nearly equal, and further 
addition of operators (via gradient-informed selection) to the ansatz fails to escape 
from this convergence plateau. This phenomenon originates from the fact that ADAPT-VQE 
selects operators with the largest gradient, but in these critical regions where ADAPT-VQE energy 
starts to converge near one or more excited states, a gradient trough emerges. Consequently, the 
operator gradients decay exponentially, leaving the ansatz unable to escape from the plateau.
This observation indicates that beyond a critical 
bond length, the SD operator pool lacks sufficient flexibility to recover the exact 
ground state, even though the energy remains very close to the FCI GS value. 
When generalized single and double excitations are included in the operator pool, ADAPT-VQE (gSD) 
still encounters the same convergence plateau. However, in this scenario, the algorithm
ultimately escapes from that plateau, recovering a state with almost unit overlap with 
the ground state. However, its overlap with the first excited state $^1A_{1g}$ remains appreciably high rather than 
vanishing, highlighting contamination from the excited state manifold.

In contrast, COMPASS-PRO (Warm), upon convergence, consistently achieves nearly unit overlap 
with the ground $^1A_{1g}$ state and negligible overlap with the first excited $^1A_{1g}$ state across all 
bond lengths considered. Moreover, it is noteworthy that COMPASS-PRO exhibits a better initial 
direction compared to both ADAPT-VQE (SD) and ADAPT-VQE (gSD). These observations suggest that 
in the challenging bond dissociation regime, ADAPT-VQE fails to reliably recover the true GS, whereas COMPASS-PRO (Warm) is highly robust to overcome the convergence plateau to reach the nearly exact GS.

\section{Conclusions and Future Outlooks}
The ordering of operators in the disentangled or pseudo-trotterized ansatz is 
largely heuristic in nature, yet this choice strongly influences not only the 
final accuracy but also, and more importantly, the nature of the convergence 
landscape. Given the limited and error-prone nature of today’s quantum 
devices, it becomes crucial to design ansatze that remain compact but expressive 
enough to recover the correct molecular energetics while simultaneously being able 
to follow an efficient optimization trajectory. In this regard, our proposed 
strategy, COMPASS-PRO, offers a near-optimal but highly practical pathway to 
approach the true ground state of a molecular Hamiltonian. 

The strength of COMPASS-PRO lies in its ability to balance precision with quantum 
resource requirements. Firstly, in the constructed operator block, essential 
higher-rank excitation effects are effectively incorporated using only lower-rank 
two-body operators, which have the ability to capture the dominant correlation 
contributions. Secondly, its progressive construction naturally provides the 
flexibility to control this trade-off, leaving the balance in the hands of the 
user. Although the ansatz-building process requires additional quantum measurements, 
this effort introduces robustness into the energy landscape and ensures
burrowing of the optimization trajectory to near the exact value. In 
practice, this means that even if the variational optimization falls into local 
traps, the protocol effectively deepens such minima and guides the optimization 
toward the true global minimum.

The COMPASS-PRO (warm) ansatz has the robustness that it can bypass local 
minima that otherwise hinders convergence. This results in a quasi-optimal 
operator ordering that yields a compact circuit with 
significantly reduced depth, without sacrificing accuracy. Remarkably, COMPASS-PRO 
is also capable of recovering the exact ground state even in the challenging 
bond-dissociation regimes where the state-of-the-art approaches like ADAPT-VQE 
fail. In addition, by replacing the standard fermionic excitation operators with qubit 
excitation operators, the circuit depth can be reduced even further, greatly 
improving the suitability of the method for near-term quantum devices.

With its compact structure, resource efficiency, and flexibility, COMPASS-PRO holds 
promise as a strong candidate for molecular simulations on NISQ hardware, 
especially when it is coupled with proper error mitigation 
techniques. Furthermore, extending this methodology to access excited-
state energetics will be an exciting direction for future development.

\section{Acknowledgment}
DM thanks Prime Minister's Research Fellowship (PMRF), Government of India for his research fellowship. R.M. acknowledges the financial support from the Anusandhan National Research Foundation 
(ANRF) (erstwhile SERB), Government of India (Grant Number: MTR/2023/001306).

\section*{Author Declarations}
\subsection*{Conflict of Interest:}
The authors have no conflict of interests to disclose.

\subsection*{DATA AVAILABILITY}
The numerical data that support the findings of this study are
available from the corresponding author upon 
reasonable request.

\bibliography{literature}

\end{document}